\documentclass[preprint,amsmath,amssymb,prd,superscriptaddress]{revtex4}
\usepackage{graphicx}% Include figure files
\newcommand{\p}{\partial}
\newcommand{\pslash}{p\kern-1ex /}
\newcommand{\lslash}{l\kern-1ex /}
\newcommand{\kslash}{k\kern-1ex /}
\newcommand{\dslash}{\p\kern-1.2ex /}
\newcommand{\Dslash}{{\cal D}\kern-1.5ex /}
\newcommand{\Tr}{{\rm Tr}}
\newcommand{\tr}{{\rm tr}}
\newcommand{\re}{{\rm Re}}

\newcommand{\sign}{{\rm sign}}
\def\Id{\mbox{1\hspace{-1.2mm}I} }

\newcommand{\bea}{\begin{eqnarray}}
\newcommand{\eea}{\end{eqnarray}}

\newcommand{\nn}{\nonumber\\}

\newcommand{\BAN}{\begin{eqnarray*}}
\newcommand{\EAN}{\end{eqnarray*}}
\begin{document}

\newcommand{\NTU}{
  Department of Physics,
  National Taiwan University, Taipei~10617, Taiwan
}

\newcommand{\CQSE}{
  Center for Quantum Science and Engineering,
  National Taiwan University, Taipei~10617, Taiwan
}

\newcommand{\CTS}{
  Center for Theoretical Sciences,
  National Taiwan University, Taipei~10617, Taiwan
}

\preprint{NTUTH-12-505A}

\title{Chiral Symmetry and the Residual Mass in Lattice QCD\\
       with the Optimal Domain-Wall Fermion }

\author{Yu-Chih~Chen}
\affiliation{\NTU}

\author{Ting-Wai~Chiu}
\affiliation{\NTU}
\affiliation{\CQSE}
\affiliation{\CTS}

\collaboration{for the TWQCD Collaboration}
\noaffiliation


\begin{abstract}

We derive the axial Ward identity for lattice QCD with domain-wall fermions,
and from which we obtain a formula for the residual mass 
(\ref{eq:mres_Dc})-(\ref{eq:Mres_Dc}), that can be used to measure the chiral symmetry 
breaking due to the finite extension $ N_s $ in the fifth dimension. 
Furthermore, we obtain an upper bound for the residual mass in lattice QCD 
with the optimal domain-wall fermion. 
%which is an exponentially decreasing 
%function of $ N_s $, for any lattice size, at zero or finite temperatures.   

\end{abstract}

\maketitle

\section{Introduction}

The chiral symmetry of massless fermion field plays an important role
in particle physics. It forbids the additive mass renormalization
which causes the fine-tuning problem associated with the scalar field.
In QCD, the chiral symmetry [$ SU_L(N_f) \times SU_R(N_f) $] 
of $ N_f $ massless quarks is spontaneously broken to
$ SU_V(N_f) $, due to the strong interaction between
quarks and gluons. This gives the (nearly) massless Goldstone
bosons (pions) and their specific interactions.
To investigate the spontaneously chiral symmetry breaking
(or hadronic physics) in QCD, it requires nonperturbative methods.
So far, lattice QCD is the most promising approach.
However, in lattice QCD, formulating lattice fermion
with exact chiral symmetry at finite lattice spacing is rather nontrivial,
which is realized by the domain-wall fermion (DWF) on the (4+1)-dimensional
lattice \cite{Kaplan:1992bt}, and the overlap fermion on the 4-dimensional lattice
\cite{Neuberger:1998fp}.

For lattice QCD with DWF, in practice, one can only use a finite number
$ N_s $ of sites in the fifth dimension. Thus the chiral symmetry of
the massless quark fields is broken, and the emergent question is whether 
the chiral symmetry is preserved optimally.
The answer is negative since the effective 4-dimensional
Dirac operator of the conventional DWF corresponds to the overlap Dirac operator 
with the polar approximation of the sign function of $ H $.

In 2002, one of us (TWC) constructed the optimal domain-wall fermion (ODWF)
\cite{Chiu:2002ir} such that the effective 4D lattice Dirac operator
attains the mathematically optimal chiral symmetry for any finite $ N_s $,
exponentially-local for sufficiently smooth gauge backgrounds \cite{Chiu:2002kj},
and independent of the lattice spacing in the fifth dimension.
The basic idea of ODWF is to construct a set of analytical weights,
$ \{ \omega_s, s = 1, \cdots, N_s \} $,
one for each layer in the fifth dimension, such that
the chiral symmetry breaking due to finite $ N_s $
can be reduced to the minimum.
The 4-dimensional effective Dirac operator of massless ODWF is
%\begin{widetext}
\BAN
%\label{eq:odwf_4d}
\begin{aligned}
D &=\frac{1}{2r} [1+ \gamma_5 S_{opt}(H) ], \\
S_{opt}(H) &= \frac{1-\prod_{s=1}^{N_s} T_s}{1 + \prod_{s=1}^{N_s} T_s}, \quad
T_s = \frac{1-\omega_s H}{1+\omega_s H},
\end{aligned}
\EAN
%\end{widetext}
which is exactly equal to the Zolotarev optimal rational approximation
of the overlap Dirac operator. That is,
$ S_{opt}(H) = H R_Z(H) $, where $ R_Z(H)$ is the optimal
rational approximation of $ (H^2)^{-1/2} $
\cite{Akhiezer:1992, Chiu:2002eh}.

However, in the original ODWF formulation \cite{Chiu:2002ir},
the valence quark propagator cannot be expressed in terms
of the correlation function of the quark fields defined
in terms of the boundary modes, unlike the conventional domain-wall fermion.
In 2003, one of us (TWC) solved this problem by introduced two transparent layers
with $ \omega_s = 0 $ \cite{Chiu:2003ir}, as boundary layers appending to the
original action of ODWF such that the quark fields defined in terms of these
two transparent layers obey the usual chiral projection rule in the continuum,
independent of the gauge fields.
Consequently any observable constructed with the quark fields manifests the
symmetries exactly as those of its counterpart in the continuum.
The salient feature of a transparent layer (with $ \omega_s = 0 $, and $ T_s = 1 $)
is that its presence does not change the effective 4D Dirac operator.

In this paper, we derive the axial Ward identity for lattice QCD with ODWF.
We find that it is necessary to extend the idea of transparent layers
introduced in Ref. \cite{Chiu:2003ir}, to add another two transparent layers
at the central region of the fifth dimension.
With these four transparent layers, the action of lattice QCD with ODWF
can be written as
\bea
\label{eq:ODWF}
%\begin{aligned}
{\cal A}_f = \sum_{s,s'=0}^{N_s+3} \sum_{x,x'}
\bar\psi_s(x)
\{ (\rho_s D_w + \Id )_{x,x'} \delta_{s,s'}
 + (\sigma_s D_w - \Id)_{x,x'}
    (P_{-} \delta_{s',s+1} + P_{+} \delta_{s',s-1}) \} \psi_{s'}(x'),
% &\equiv \sum_{s,s'=0}^{N_s+3} \sum_{x,x'} \bar\psi_s(x) {\cal D}_{x,s;x's'} \psi_{s'}(x'),
%\end{aligned}
\eea
with boundary conditions
\bea
\begin{aligned}
P_{+} \psi(x,-1) &= - r m_q P_{+} \psi(x,N_s+3), \\
P_{-} \psi(x,N_s+4) &= - r m_q P_{-} \psi(x,0),
\label{eq:bc}
\end{aligned}
\eea
where
$ P_{\pm} = (1 \pm \gamma_5)/2 $,
$ D_w $ is the standard Wilson-Dirac operator plus a negative parameter
$ -m_0 $ ($ 0 < m_0 < 2 $),
$ m_q $ is the bare quark mass, and $ r $ is a parameter depending on
$ \{ \rho_s, \sigma_s \} $ and $ m_0 $ such that the valence
quark propagator agrees with $ (\gamma_\mu \p_\mu + m_q)^{-1} $ in the continuum limit.
The two transparent layers at the boundaries
are specified by imposing $\rho_{0}=\rho_{N_s+3}=\sigma_{0}=\sigma_{N_s+3}=0$.
The two additional transparent layers can be located at $ s = n $, and $ s=n+1 $,
where $ n = [N_s/2]$. In other words, they have
$\rho_{n}=\rho_{n+1}=\sigma_{n}=\sigma_{n+1}=0$.
In the original ODWF formulation \cite{Chiu:2002ir}, 
the nonzero $ \rho_s $ and $ \sigma_s $ are set to be $ \omega_s $ (the optimal weight).

The quark fields are defined in terms of the boundary modes
\bea
\begin{aligned}
\label{eq:q_odwf}
q(x) &= \sqrt{r} \left[ P_{-} \psi_0(x) + P_{+} \psi_{N_s+3}(x) \right], \\
\label{eq:bar_q_odwf}
\bar q(x) &= \sqrt{r} \left[ \bar\psi_{0}(x) P_{+} + \bar\psi_{N_s+3}(x) P_{-} \right].
\end{aligned}
\eea

Following the derivation given in Ref. \cite{Chiu:2003ir},
it is straightforward to show (in Section \ref{Z})
that the valence quark propagator in a gauge background is 
equal to the correlation function of the quark fields, i.e.,
\bea
\label{eq:Dcmi}
\langle q(x) \bar q(y) \rangle = (D_c + m_q)^{-1}(x,y),
\eea
where
\bea
\label{eq:Dc}
&& D_c = \frac{1}{r} \frac{ 1 + \gamma_5 S }{1 - \gamma_5 S}, \\
\label{eq:S_opt}
&& S = \frac{1-\prod_{s=0}^{N_s+3} T_s}{1 + \prod_{s=0}^{N_s+3} T_s}, \\
\label{eq:Ts}
&& T_s = \frac{1-H_s}{1+H_s}, \\
\label{eq:Hs}
&& H_s = (\rho_s + \sigma_s) H_w [ 2 + (\rho_s - \sigma_s) \gamma_5 H_w ]^{-1}, \hspace{4mm} H_w = \gamma_5 D_w.
\eea
Obviously, a transparent layer (with $ \rho_s = \sigma_s = 0 $)
does not change $ S $ and $ D_c $ since its $ T_s = 1 $.
Setting the nonzero weights $ \rho_s = c \ \omega_s + d $, and $ \sigma_s = c \ \omega_s - d $,
where $ c $ and $ d $ are constants, then $ H_s = \omega_s H $,
\bea
\label{eq:H}
H = c H_w ( 1 + d \gamma_5 H_w )^{-1}, 
\eea
and the parameter $ r $ entering the boundary conditions (\ref{eq:bc})  
is fixed to $ r = [2m_0(1-d m_0)]^{-1} $ such that $ (D_c + m_q)^{-1} $ 
in the free fermion limit agrees with $ (\gamma_\mu \p_\mu + m_q)^{-1} $ 
in the continuum limit. 
Moreover, for $ H_s = \omega_s H $, interchanging any two layers in the 
fifth dimension gives the same $ S $, since $ \{ T_s \} $ commute among themselves.

For finite $ N_s $, with the optimal weights $ \{ \omega_s \} $ given in Ref. \cite{Chiu:2002ir},
$ S $ is exactly equal to the Zolotarev optimal rational approximation of the sign function of $ H $, 
i.e., $ S = S_{opt}(H) = H R_Z(H) $, where $ R_Z(H)$ is the optimal
rational approximation of $ (H^2)^{-1/2} $ \cite{Akhiezer:1992, Chiu:2002eh}.
In the limit $ N_s \to \infty $, $ S \to H (H^2)^{-1/2} $,
and $ D_c $ becomes exactly chirally symmetric, and $ (D_c + m_q)^{-1} $ is well-defined for
nonzero $ m_q $, even though $ D_c $ is ill-defined for topologically nontrivial
gauge background \cite{Chiu:1998gp}.

In practice, only the case $ d = 0 $ gives $ H = c H_w $ (without the denominator), 
which is much easier for the projection of the low-lying eigenmodes of $ D = D_c (1+rD_c)^{-1} $ 
than other cases with $ d \ne 0 $.
Since the low-lying eigenmodes of $ D $ are vital for extracting many physical observables,
the original formulation \cite{Chiu:2002ir} with $ d = 0 $ (and $ c = 1 $) 
seems to be a good choice. 

We note in passing that setting the nonzero weights $ \rho_s = c_1 (\text{constant}) $ and 
$ \sigma_s = c_2 (\text{constant}) $ covers all variants of conventional domain-wall fermions, 
with $ S $ equal to the polar approximation of the sign function of $ H $, 
\BAN
\label{eq:S_polar}
S_{polar}(H) =  
\begin{cases}
             H \left( \frac{1}{N_s} +
                      \frac{2}{N_s} \sum_{l=1}^{n} \frac{b_l}{H^2 + d_l} \right), 
             &  \ N_s = 2n+1 \mbox{ (odd) }, \\
             H \
             \frac{2}{N_s} \sum_{l=1}^{n} \frac{b_l}{H^2 + d_l}, 
             &  \ N_s = 2n  \mbox{ (even) }, 
\end{cases}
\EAN
where
\BAN
\label{eq:polar_bd}
b_l = \sec^2 \left[ \frac{\pi}{N_s} \left( l-\frac{1}{2} \right) \right],
\hspace{4mm}
d_l = \tan^2 \left[ \frac{\pi}{N_s} \left( l-\frac{1}{2} \right) \right] \ .
\EAN
For example, setting $ \rho_s = 1 $ and $ \sigma_s = 0 $,
(\ref{eq:ODWF}) reduces to the conventional domain-wall fermion with 
$ H = H_w ( 2 + \gamma_5 H_w )^{-1} $ \cite{Shamir:1993zy}, 
and $ \rho_s = \sigma_s = 1 $ to the Borici's variant with $ H = H_w $ \cite{Borici:1999zw}, 
and $ \rho_s = c + d $ and $ \sigma_s = c - d $ to the M\"obius variant with 
$ H = c H_w ( 1 + d \gamma_5 H_w )^{-1} $ \cite{Brower:2004xi}.

\section{Axial Ward Identity}

Now we consider $ N_f $ flavors of quarks with degenerate mass $ m_q $, and
the infinitesimal flavor non-singlet transformation
\bea
\label{eq:ct}
\begin{aligned}
\delta \psi_s(x) &= i \theta_s(x) \lambda^a \psi_s(x), \\
\delta \bar\psi_s(x) &= -i\theta_s(x) \bar\psi_s(x) \lambda^a,  
\end{aligned}
\eea
where
\BAN
\theta_s(x) =
\begin{cases}
\theta(x),  &  0 \leq s \leq n \equiv \left[\frac{N_s}{2} \right], \nn
-\theta(x), &  n+1 \leq s \leq N_s+3.
\end{cases}
\EAN
Here $ \lambda^a $ is one of the flavor group generators in the fundamental representation,
and the flavor indices of $ \psi_s(x) $ and $ \bar\psi_s(x) $ are suppressed.
Under the transformation (\ref{eq:ct}), the change of the action (\ref{eq:ODWF}) consists of
the following three parts:
\bea
%\label{eq:partA}
&& \delta \sum_{s=0}^{N_s+3} \sum_{x,y} \rho_s \left[ \bar\psi_s(x) \lambda^a D_w(x,y) \psi_s(y)  \right]
= \sum_x i\theta(x) \sum_{\mu} \Delta_{\mu} \hat j^a_{\mu}(x), \nn
%\label{eq:partB}
&& \delta \sum_{s=0}^{N_s+3} [-\bar\psi_s(x) \lambda^a P_- \psi_{s+1}(x)- \bar\psi_s(x) \lambda^a P_+ \psi_{s-1}(x)]
= - \sum_x 2 i \theta(x) [ J^a_5(x,n)+ m_q \bar q(x) \lambda^a \gamma_5 q(x)], \nn
\label{eq:partC}
&& \delta \sum_{s=0}^{N_s+3} \sum_{x,y} \sigma_s \{ \bar\psi_s(x) \lambda^a D_w(x,y)
                          [ P_- \psi_{s+1}(y)  +  P_+ \psi_{s-1}(y)]\}
= \sum_x i\theta(x) \sum_{\mu} \Delta_{\mu} \hat k^a_{\mu}(x), 
\eea
where
\bea
&& \Delta_{\mu} f(x,s) \equiv f(x,s)-f(x-\mu,s), \nn
&& \hat j^a_{\mu}(x) \equiv \sum_{s=1}^{N_s+2} \sign \left(n-s+\frac{1}{2} \right) j^a_{\mu}(x,s), \nn
&& j^a_{\mu}(x,s) = \frac{\rho_s}{2} \left[\bar\psi_s(x) \lambda^a (1-\gamma_{\mu})U_{\mu}(x)\psi_s(x+\mu)
                          -\bar\psi_s(x+\mu) \lambda^a (1+\gamma_{\mu})U^{\dag}_{\mu}(x)\psi_s(x) \right],  \nn
\label{eq:def_J5}
&& J^a_5(x,n) = -\bar\psi_n(x) \lambda^a P_- \psi_{n+1}(x) + \bar\psi_{n+1}(x) \lambda^a P_+ \psi_{n}(x), \\
&& \hat k_{\mu}^a(x) \equiv {\hat k_{\mu}}^{a+}(x) + {\hat k_{\mu}}^{a-}(x), \nn
&& \hat k_{\mu}^{a\pm}(x) \equiv \sum_{s=1}^{N_s+2} \sign \left(n-s+\frac{1}{2} \right) k_{\mu}^{a\pm}(x,s),  \nn
&& k_{\mu}^{a\pm}(x,s) =
    \frac{\sigma_s}{2} \left[\bar\psi_s(x) \lambda^a (1-\gamma_{\mu})U_{\mu}(x)P_{\pm}\psi_{s\mp1}(x+\mu)
        -\bar\psi_s(x+\mu) \lambda^a (1+\gamma_{\mu})U^{\dag}_{\mu}(x)P_{\pm}\psi_{s\mp1}(x) \right]. \nonumber
\eea

Now the role of the two transparent layers at $s=n$ and $s=n+1$ becomes obvious.
If we want to keep $ J_5^a $ (\ref{eq:def_J5}) not depending on $ D_w $ 
(similar to the $ J_5^a $ in the conventional DWF) and to express (\ref{eq:partC}) 
in terms of the divergence of a 4-current, then it is inevitable to introduce 
two transparent layers in the central region of the 5th dimension. 
This can be seen as follows.
For $ 1 \le s \le n-1 $, or $ n+2 \le s \le N_s + 2 $, we have
\BAN
\delta\sum_{x,y} \sigma_s  \bar\psi_s(x) \lambda^a D_w(x,y)P_- \psi_{s+1}(y)
= \mp i \sum_{x,y}\bar\psi_s(x)\lambda^a\sigma_s[\theta(x)D_w(x,y)-D_w(x,y)\theta(y)]P_- \psi_{s+1}(y),
\EAN
which can be written in the form of $\sum_{x}\theta(x)\Delta_{\mu}J_{\mu}$.
However, at $s=n$, it gives
\BAN
\delta\sum_{x,y} \sigma_n  \bar\psi_n(x) \lambda^a D_w(x,y)P_- \psi_{n+1}(y)
= -i \sum_{x,y}\bar\psi_s(x)\lambda^a\sigma_n[\theta(x)D_w(x,y)+D_w(x,y)\theta(y)]P_- \psi_{n+1}(y), 
\EAN
which cannot be expressed in terms of the divergence of a 4-current unless $ \sigma_n = 0 $.
Similarly, we also set $\sigma_{n+1}=0$. Furthermore, for consistency,
we must also set $ \rho_{n} = \rho_{n+1} = 0 $ such that $ T_n = T_{n+1} = 1 $.

For any observable $ {\cal O} $, the variation of its vacuum expectation value with respect to
(\ref{eq:ct}) must vanish, i.e., $ \delta^a \langle {\cal O} \rangle = 0$, which
gives the axial Ward identity
\bea
\label{eq:AWI_O_Nf}
\sum_{\mu}\Delta_{\mu}\langle J_{\mu}^a(x){\cal O}(y) \rangle
= 2m_q \langle \bar q(x) \lambda^a\gamma_5  q(x){\cal O}(y)\rangle
 + 2 \langle J_5^a(x,n){\cal O}(y)\rangle + i \langle \delta^a {\cal O}(y)\rangle.
\eea
where $J^a_{\mu}(x) \equiv \hat k^a_{\mu}(x)+\hat j^a_{\mu}(x)$.
As $N_s \to \infty$, the anomalous term $\langle J_5^a(x,n){\cal O}(y)\rangle$ vanishes 
if ${\cal O}(y)$ only involves the quark fields, 
following the same argument given in Ref. \cite{Furman:1994ky}.

After summing over all sites $x$, the LHS of (\ref{eq:AWI_O_Nf}) vanishes, and its RHS gives
\bea
\label{eq:AWI_O_sum_x}
\begin{aligned}
 -i \sum_{x}\langle \delta^a {\cal O}(y) \rangle 
= 2 m_q \sum_{x}\langle \bar q(x) \lambda^a\gamma_5 q(x) {\cal O}(y) \rangle
 + 2 \sum_{x}\langle J_5^a(x,n) {\cal O}(y) \rangle.
\end{aligned}
\eea
Thus, the effect of chiral symmetry breaking due to finite $ N_s $ can be
regarded as an additive mass to the bare quark mass $ m_q $, the so-called residual mass
\bea
\label{eq:def_mres_O}
m_{res}[{\cal O}(y)] = \frac{\sum_{x}\langle J_5^a(x,n) {\cal O}(y) \rangle}
                            {\sum_{x}\langle \bar q(x) \lambda^a\gamma_5 q(x) {\cal O}(y) \rangle}, 
\eea
which serves as a measure of the chiral symmetry breaking due to finite $ N_s $.
In the limit $ N_s \to \infty $, $ S(H) = H/\sqrt{H^2} $ and $ m_{res} = 0 $.
Obviously, in a gauge background, the residual mass (\ref{eq:def_mres_O}) 
depends on the observable $ {\cal O} $ as well as its location $ y $. 
Thus it is necessary to take into account of the residual mass at all locations. 
This can be accomplished by summing over all lattice sites $ y $ 
in the axial Ward identity (\ref{eq:AWI_O_sum_x}) to obtain the global residual mass
\bea
\label{eq:def_Mres_O}
M_{res}[{\cal O}] = \frac{\sum_{x,y}\langle J_5^a(x,n) {\cal O}(y) \rangle}
                          {\sum_{x,y}\langle \bar q(x) \lambda^a\gamma_5 q(x) {\cal O}(y) \rangle}. 
\eea
For $ {\cal O}(y)= \bar q(y) \lambda^b \gamma_5 q(y) $, (\ref{eq:def_mres_O}) and (\ref{eq:def_Mres_O}) become 
\bea
\label{eq:def_mres}
m_{res}(y) &=& \frac{\sum_{x}\langle J_5^a(x,n)\bar q(y) \lambda^b\gamma_5  q(y)\rangle}
                    {\sum_{x}\langle \bar q(x) \lambda^a\gamma_5 q(x)\bar q(y) \lambda^b\gamma_5 q(y)\rangle}, \\
\label{eq:def_Mres}
M_{res} &=& \frac{\sum_{x,y}\langle J_5^a(x,n)\bar q(y) \lambda^b\gamma_5  q(y)\rangle}
               {\sum_{x,y}\langle \bar q(x) \lambda^a\gamma_5 q(x)\bar q(y) \lambda^b\gamma_5 q(y)\rangle},  
\eea
which are usually used as a measure of the chiral symmetry breaking due to finite $ N_s $.
In the following, we will restrict our discussions to the residual mass (\ref{eq:def_mres}), 
and the global residual mass (\ref{eq:def_Mres}).

\section{Generating functional for $n$-point Green's function}
\label{Z}

In order to express the residual mass (\ref{eq:def_mres}) in terms of the quark propagator, 
we first derive the generating functional for the $n$-point Green's function of the fermion fields for 
lattice QCD with ODWF.
With four transparent layers, the action (\ref{eq:ODWF}) can be rewritten as
\bea
\label{eq:ODWF_A}
{\cal A}_f
&=& \sum_{s=0}^{N_s+3} \sum_{x,x'}\bar\psi_s(x)\gamma_5
\{ (\rho_s H_wP_{+}+ \rho_s H_wP_{-} +P_{+}-P_{-} )_{x,x'} \psi_{s}(x') \nonumber \\
& & \hspace{10mm} +(\sigma_s H_wP_{-} +\sigma_s H_wP_{+}+P_{-}-P_{+})_{x,x'}(P_{-}\psi_{s+1}(x') + P_{+}\psi_{s-1}(x')) \} \nonumber\\
&=& \sum_{s=0}^{N_s+3} \sum_{x,x'}\bar\psi_s(x)\gamma_5
    \{ ( \rho_s H_wP_{-}+\sigma_s H_wP_{+}-1)_{x,x'} [P_{+}\psi_{s-1}(x')+P_{-} \psi_{s}(x')] \nonumber\\
& & \hspace{10mm} +( \rho_s H_wP_{+}+\sigma_s H_wP_{-}+1)_{x,x'} [P_+\psi_{s}(x')+P_{-}\psi_{s+1}(x')] \} \nonumber\\
&=& \sum_{s=0}^{N_s+3} \sum_{x,x'}\bar\psi_s(x)\gamma_5
    \{ Q^s_{-}(x,x') [P_{+}\psi_{s-1}(x')+P_{-} \psi_{s}(x')]+Q^s_{+}(x,x') [P_+\psi_{s}(x')+P_{-}\psi_{s+1}(x')]\}, \nonumber
\eea
where
\bea
Q^s_{\pm} \equiv \rho_s H_w P_{\pm}+\sigma_s H_w P_{\mp} \pm 1.
\eea
Defining
\bea
\label{eq:def_eta}
&&\eta_s \equiv (P_- \delta_{s',s} + P_+ \delta_{s',s-1} ) \psi_{s'} 
 \Leftrightarrow  \psi_s = (P_- \delta_{s',s} + P_+ \delta_{s',s+1} ) \eta_{s'}, \\
\label{eq:def_etabar}
&&\bar\eta_s \equiv \bar\psi_{s}\gamma_5 Q^s_-  \Leftrightarrow  \bar\psi_s = \bar\eta_{s} (Q^s_-)^{-1} \gamma_5, \\
\label{eq:def_tran_matrix}
&&T_s \equiv -(Q^s_+)^{-1} Q^s_-= \frac{1-H_s}{1+H_s}, \hspace{4mm} 
H_s = (\rho_s + \sigma_s) H_w [ 2 + (\rho_s - \sigma_s) \gamma_5 H_w ]^{-1}, 
\eea
then the action (\ref{eq:ODWF}) can be expressed in terms of ${\eta,\bar\eta}$ fields
\bea
\label{eq:ODWF_in_eta}
{\cal A}_f &=& \bar\eta_0 (P_- - rm_q P_+)\eta_0 - \bar\eta_0 \eta_0
                +\sum_{s=1}^{N_s+2} [\bar\eta_s\eta_s - \bar\eta_s T^{-1}_s \eta_{s+1}]\nonumber\\
            & & +\bar\eta_{N_s+3}\eta_{N_s+3}  -  \bar\eta_{N_s+3}( P_+ - r m_q P_- )\eta_0,  
\eea
where the space-time indices have been suppressed.

In order to evaluate the Green's function of the fermion fields in the expression of the residual mass, 
we need to add the following external source terms to (\ref{eq:ODWF_in_eta})  
\BAN
%\label{eq:external_source_anomaly}
&&\bar\eta_{n}J_{n} +\bar J_{n+1}\eta_{n+1}+\bar\eta_{n+1}J_{n+1}, \\
%\label{eq:external_source_quark}
&& \bar J_q q + \bar q J_q = \bar J \eta_0 - \bar\eta_0P_+ J +\bar\eta_{N_s+3}P_- J,  
\EAN
where 
\BAN
&& J \equiv \sqrt{r} J_q, \\
&& \bar J \equiv \sqrt{r} \bar J_q.
\EAN 
Then the generating functional for $n$-point Green's function is defined as
\bea
\label{eq:generating_functional}
Z[J_q, \bar J_q, J_n, J_{n+1}, \bar J_{n+1}]={\cal J}\int[d\bar\eta][d\eta]e^{-S_J}, 
\eea
where
\bea
\label{eq:action_generating_functional}
{S}_J = {\cal A}_f
            -\bar J \eta_0 + \bar\eta_0 P_+ J
            -\bar\eta_{n}J_{n} - \bar J_{n+1}\eta_{n+1} - \bar\eta_{n+1}J_{n+1}
            -\bar\eta_{N_s+3} P_- J, 
\eea
and ${\cal J}$ is the Jacobian of the transformation,
\bea
\label{eq:Jacobian}
{\cal J} = \prod_{s=0}^{N_s+3} \det(\rho_s H_w P_{-}+\sigma_s H_w P_{+}-1).
\eea
Now using the Grassman integration formula
\BAN
%\label{eq:G_integral}
\int d \bar\chi d \chi \ e^{-\bar\chi M \chi + \bar v \chi + \bar \chi v}
= e^{\bar v M^{-1} v} \det M, 
\EAN
and integrating $(\eta_s, \bar\eta_s)$ successively from $s=N_s+3$ to $s=1$, (\ref{eq:generating_functional}) becomes
\bea
\label{eq:integral_eta0}
&& {\cal J}\int[d\bar\eta_0][d\eta_0]
   \exp \left\{ \bar\eta_0 \left[(P_- - rm_q P_+)-\prod_{s=1}^{N_s+2}T_s^{-1}(P_+ -rm_q P_- ) \right] \eta_0 \right. \nonumber\\
&& -\eta_0 \left[ \left(\prod_{s=1}^{N_s+2}T_s^{-1}P_- - P_+ \right)J
                  +\prod_{s=1}^{n}T_s^{-1}J_{n} + \prod_{s=1}^{n}T_s^{-1}J_{n+1} \right] \nonumber\\
&&  -\left[ \bar J +\bar J_{n+1}\prod_{s=n+1}^{N_s+2}T_s^{-1}(P_+ -rm_q P_- ) \right] \eta_0
    \left.   -\bar J_{n+1}\prod_{s=n+1}^{N_s+2}T_s^{-1}P_-J-\bar J_{n+1}J_{n+1} \right\}. 
\eea
Finally integrating $(\eta_0, \bar\eta_0)$ of (\ref{eq:integral_eta0}), we obtain the generating functional 
\bea
\label{eq:final_result_generating_functional}
&& Z[J_q, \bar J_q, J_n, J_{n+1}, \bar J_{n+1}] \nn
&=&{\cal J} \det \left[ (P_- - rm_q P_+)-\prod_{s=1}^{N_s+2}T_s^{-1}(P_+ -rm_q P_- ) \right]
  \exp\left\{ \bar J_{n+1}\prod_{s=n+1}^{N_s+2}T_s^{-1}P_-J+\bar J_{n+1}J_{n+1} + \right.  \nonumber\\
&&    + \left[ \bar J +\bar J_{n+1}\prod_{s=n+1}^{N_s+2}T_s^{-1}( P_+ -rm_q P_- ) \right] \cdot
                \left[ (P_- - rm_q P_+)-\prod_{s=1}^{N_s+2}T_s^{-1}(P_+ -rm_q P_- ) \right]^{-1} \nonumber\\
          & & \cdot \left[(\prod_{s=1}^{N_s+2}T_s^{-1}P_- - P_+ )J
         \left.            +\prod_{s=1}^{n}T_s^{-1}J_{n} + \prod_{s=1}^{n}T_s^{-1}J_{n+1} \right] \right\}  \nn
&=& K \det[r(D_c+m_q)]
     \exp\Big\{ \bar J_{n+1}T_U^{-1}P_-J+\bar J_{n+1}J_{n+1} + \nonumber\\
&&  \hspace{10mm}    + \left[ \bar J +\bar J_{n+1}T_U^{-1}(P_+ -rm_qP_- ) \right]
                        r^{-1}(D_c+m_q)^{-1}
                       \left[J+\widehat{T}^{-1}J_{n} + \widehat{T}^{-1}J_{n+1} \right] \Big\}, 
\eea
where we have used the identity
\BAN
\left(-P_+ + \prod_{s=1}^{N_s+2}T_s^{-1}P_- \right)^{-1}\left( P_- -\prod_{s=1}^{N_s+2}T_s^{-1}P_+  \right) 
= \frac{1+\gamma_5 S}{1-\gamma_5 S} = rD_c, 
\EAN
and defined
\BAN
&& T_L^{-1} \equiv  \prod_{s=1}^{n}T_s^{-1} \hspace{4mm},  \\ 
&& T_U^{-1} \equiv  \prod_{s=n+1}^{N_s+2}T_s^{-1} \hspace{4mm}, \\
&& T^{-1}  \equiv \prod_{s=1}^{N_s+2}T_s^{-1} = T_L^{-1}T_U^{-1}, \\
&& \widehat{T}^{-1} \equiv \left( -P_+ + T^{-1}P_- \right)^{-1}T_L^{-1}, \\
&& K  \equiv {\cal J} \det\left[ -P_+ + T^{-1}P_- \right]. 
\EAN
Equation (\ref{eq:final_result_generating_functional}) is one of the main results of this paper.

With the generating functional (\ref{eq:final_result_generating_functional}), we obtain the propagators in a gauge background 
as follows.

\noindent (I) The valence quark propagator 
\bea
\label{eq:quark_propagator}
\langle q(x)\bar q(y)\rangle = \left. -\frac{1}{Z}\frac{\delta^2 Z} {\delta \bar J_q(x) \delta J_q(y)} \right|_0 
                             = (D_c+m_q)^{-1}(x,y), 
\eea
where the subscript 0 in the functional derivative denotes setting all $J$'s to zero after differentiation. 

\noindent (II) The mixed correlator of the first kind
\bea
\label{eq:q_etabar_n}
\langle q(x)\bar\eta_n(y)\rangle
    &=& \left. -\frac{1}{Z}\frac{\delta^2 Z}{\delta \bar J_q(x) \delta J_n(y)} \right|_0  \nonumber\\
    &=& \frac{1}{\sqrt{r}}(D_c+m_q)^{-1} \left( -P_+ + T^{-1}P_- \right)^{-1}T_L^{-1} \nonumber\\
    &=& -\frac{1}{\sqrt{r}}D^{-1}(m_q)\gamma_5\frac{T_L^{-1}}{T^{-1}+1}, 
\eea
where
\bea
\label{eq:DcmI}
D^{-1}(m_q) = (1+rD_c)(D_c + m_q)^{-1} = r + ( 1- r m_q) ( D_c + m_q)^{-1}, 
\eea
the sea quark propagator.

\noindent (III) The mixed correlator of the second kind
\bea
\label{eq:q_etabar_n1}
\langle q(x)\bar\eta_{n+1}(y)\rangle
    &=& \left. -\frac{1}{Z}\frac{\delta^2 Z}
                     {\delta \bar J_q(x) \delta J_{n+1}(y)} \right|_0  \nonumber\\
    &=& \frac{1}{\sqrt{r}}(D_c+m_q)^{-1}\left( -P_+ + T^{-1}P_- \right)^{-1}T_L^{-1}\nonumber\\
    &=& -\frac{1}{\sqrt{r}}D^{-1}(m_q)\gamma_5\frac{T_L^{-1}}{T^{-1}+1}
     = \langle q(x)\bar\eta_n(y)\rangle.
\eea

\noindent (IV) The mixed correlator of the third kind
\bea
\label{eq:eta_n1_qbar}
\langle \eta_{n+1}(x) \bar q(y)\rangle
&=& \left. -\frac{1}{Z}\frac{\delta^2 Z}{\delta \bar J_{n+1}(x) \delta J_q(y)} \right|_0 \nonumber\\
&=& T_U^{-1}(-rm_qP_- +P_+)(D_c+m_q)^{-1} \frac{1}{\sqrt{r}}+T_U^{-1}P_- \sqrt{r} \nonumber\\
&=& T_U^{-1}\frac{1}{2\sqrt{r}}\left(1+\frac{1+rm_q}{1-rm_q}\gamma_5 \right) D^{-1}(m_q)-\frac{\sqrt{r}}{1-rm_q}T_U^{-1}\gamma_5.
\eea

For completeness, we also consider the generating functional for $n$-point Green's function of fermion fields in full QCD with ODWF 
(satisfying the normalization condition $ Z[0] = 1 $)
\bea
\label{eq:ZW}
Z[J_q, \bar J_q, J_n, J_{n+1}, \bar J_{n+1}] = 
\frac{ \int e^{ -{\cal A}_g -{\cal A}_f -{\cal A}_{PV}
                           -\bar J \eta_0 + \bar\eta_0 P_+ J
                           -\bar\eta_{n}J_{n} - \bar J_{n+1}\eta_{n+1} - \bar\eta_{n+1}J_{n+1}
                           -\bar\eta_{N_s+3} P_- J } 
} 
{ \int e^{-{\cal A}_g -{\cal A}_f -{\cal A}_{PV}} },  
\eea
where
$ \int \equiv \int [dU][d\psi][d\bar\psi][d\phi][d\bar\phi] $,
$ {\cal A}_g $ is the gauge action, and
$ {\cal A}_{PV} $ is the action of the Pauli-Villars fields 
$ \{ \bar \phi_s, \phi_s \} $ with $ m_q = 1/r $, i.e.,  
\BAN
%\label{eq:A_PV}
%\begin{aligned}
{\cal A}_{PV} = \sum_{s,s'=0}^{N_s+3} \sum_{x,x'} \bar\phi_s(x) \{ (\rho_s D_w + \Id )_{x,x'} \delta_{s,s'}
               + (\sigma_s D_w - \Id)_{x,x'} (P_{-} \delta_{s',s+1} + P_{+} \delta_{s',s-1}) \} \phi_{s'}(x'),
% &\equiv \sum_{s,s'=0}^{N_s+3} \sum_{x,x'} \bar\psi_s(x) {\cal D}_{x,s;x's'} \psi_{s'}(x'),
%\end{aligned}
\EAN
with boundary conditions
\BAN
\begin{aligned}
P_{+} \phi(x,-1) &= - P_{+} \phi(x,N_s+3), \\
P_{-} \phi(x,N_s+4) &= - P_{-} \phi(x,0).
%\label{eq:bc}
\end{aligned}
\EAN

Since the integrals over the fermion fields have been done, 
we proceed to evaluate the integrals over the Pauli-Villars fields in
(\ref{eq:ZW}). Using the Gaussian integration formula for the
boson fields, and following the procedures similar to
above for the fermion fields, we obtain
\bea
\label{eq:int_PV}
\int [d \bar\phi][d \phi] e^{-{\cal A}_{PV}}
= \pi^{N_s+4} K^{-1}  \det (1+r D_c)^{-1}.
\eea

Substituting (\ref{eq:final_result_generating_functional}), and (\ref{eq:int_PV})
into (\ref{eq:ZW}), we have
\bea
\label{eq:ZW_odwf}
& &  Z[J_q, \bar J_q, J_n, J_{n+1}, \bar J_{n+1}] = \frac{1}{ \int [dU] e^{-{\cal A}_g} \det D(m_q) } \times \nn
& & \hspace{8mm} \int [dU] e^{-{\cal A}_g} \det D(m_q) 
     \exp\Big\{ \bar J_{n+1}T_U^{-1}P_-J+\bar J_{n+1}J_{n+1} + \nn
&&  \hspace{10mm}    + \left[ \bar J +\bar J_{n+1}T_U^{-1}(P_+ -rm_qP_- ) \right]
                        r^{-1}(D_c+m_q)^{-1}
                       \left[J+\widehat{T}^{-1}J_{n} + \widehat{T}^{-1}J_{n+1} \right] \Big\}, 
\eea
where
\bea
\begin{aligned}
D(m_q) &= (D_c + m_q)(1 + r D_c)^{-1} \\
       &= m_q + \frac{1}{2} \left(\frac{1}{r}- m_q \right) (1+\gamma_5 S),
\end{aligned}
\eea
the effective 4D lattice Dirac operator, with $ D_c $ and $ S $ defined in Eqs. (\ref{eq:Dc})-(\ref{eq:Hs}).  
Setting the nonzero weights $ \rho_s = c \ \omega_s + d $, and $ \sigma_s = c \ \omega_s - d $,
where $ c $ and $ d $ are constants, then $ H_s = \omega_s H $ with 
$ H = c H_w ( 1 + d \gamma_5 H_w )^{-1} $.
For finite $ N_s $, with the optimal weights $ \{ \omega_s \} $ given in Ref. \cite{Chiu:2002ir},
$ S $ is exactly equal to the Zolotarev optimal rational approximation of the sign function of $ H $, 
i.e., $ S = S_{opt}(H) = H R_Z(H) $, where $ R_Z(H)$ is the optimal
rational approximation of $ (H^2)^{-1/2} $ \cite{Akhiezer:1992, Chiu:2002eh}.
In the limit $ N_s \to \infty $, $ S \to H (H^2)^{-1/2} $, and $ D(0) $ is exactly equal to
the overalap Dirac operator, satisfying the Ginsparg-Wilson relation \cite{Ginsparg:1981bj}
\BAN
D(0) \gamma_5 + \gamma_5 D(0) = 2 r D(0) \gamma_5 D(0).
\EAN

\section{A Formula for the Residual Mass}
\label{formula_mres}

Now we are ready to derive a formula for the residual mass, in terms of
the quark propagator in a gauge background. 
The denominator of (\ref{eq:def_mres}) can be evaluated as
\bea
\label{eq:denominator_mres}
& &  \sum_{x}\langle \bar q(x) \lambda^a\gamma_5  q(x)\bar q(y) \lambda^b\gamma_5  q(y)\rangle \nonumber\\
&=& - \tr_F(\lambda^a\lambda^b) \tr_{DC} \{ [(D_c + m_q)^{-1}]^\dagger  (D_c + m_q)^{-1} \}(y,y),
\eea
where the subscript $ F $ denotes the flavor space, and the subscript $ DC $ the Dirac and color spaces. In the following,  
the subscripts $ F $ and $ DC $ will be suppressed.

Using Eqs. (\ref{eq:def_J5}), (\ref{eq:def_eta}), (\ref{eq:def_etabar}),  
and (\ref{eq:q_etabar_n})-(\ref{eq:eta_n1_qbar}), we evaluate the numerator of (\ref{eq:def_mres}) as  
\bea
\label{eq:numerator_mres}
& & \sum_{x}\langle J_5^a(x,n)\bar q(y) \lambda^b\gamma_5  q(y)\rangle \nn
&=& \tr(\lambda^a\lambda^b) \left\{\sum_{x}\tr[\langle q(y)\bar \psi_n(x)\rangle P_- \langle \psi_{n+1}(x)\bar q(y)\rangle\gamma_5]
  - \sum_{x}\tr[\langle q(y)\bar \psi_{n+1}(x)\rangle P_+ \langle \psi_n(x)\bar q(y)\rangle\gamma_5] \right\} \nn
&=& \tr(\lambda^a\lambda^b) \left\{-\sum_{x}\tr[\langle q(y)\bar \eta_n(x)\rangle \gamma_5P_- \langle \eta_{n+1}(x)\bar q(y)\rangle\gamma_5]
  + \sum_{x}\tr[\langle q(y)\bar \eta_{n+1}(x)\rangle \gamma_5P_+ \langle \eta_{n+1}(x)\bar q(y)\rangle\gamma_5] \right\} \nn
&=& \tr(\lambda^a\lambda^b) \sum_{x}\tr[\langle q(y)\bar \eta_n(x)\rangle \langle \eta_{n+1}(x)\bar q(y)\rangle \gamma_5] \nn
%&=& \tr(\lambda^a\lambda^b)\frac{1}{(1-rm_q)}\tr[\gamma_5D^{-1}(m_q)(r D-P_-)\gamma_5](y,y) \nn
%& & -\tr(\lambda^a\lambda^b) \frac{1}{2r(1-rm_q)}
%     \tr\left[\gamma_5D^{-1}(m_q)(rD-P_-)\{1-rm_q+(1+rm_q)\gamma_5\}D^{-1}(m_q)  \right](y,y) \nn
&=& \tr(\lambda^a\lambda^b)\frac{1}{r}\tr[\gamma_5D^{-1}(m_q)(rD - P_-)\gamma_5(r D - P_+)D^{-1}(m_q)](y,y) \nn
&=& -\tr(\lambda^a\lambda^b) \frac{1}{4r} \tr\{ [D^{-1}(m_q)]^{\dagger} (1-S^2) D^{-1}(m_q)](y,y).
\eea
where $ D = D(0) = D_c ( 1 + r D_c )^{-1} $.
%\bea
%&& D = D(0) = D_c ( 1 + r D_c )^{-1}, \\ 
%&& \frac{T^{-1}}{T^{-1}+1} = \frac{1}{2}(1+S) = \gamma_5 (r D) + P_-. 
%\eea
Using (\ref{eq:denominator_mres}) and (\ref{eq:numerator_mres}), we can rewrite (\ref{eq:def_mres}) as   
\bea
\label{eq:mres_delta}
m_{res}(y) = \frac{1}{4r} \frac{\tr\{ [D^{-1}(m_q)]^{\dagger}(1-S^2) D^{-1}(m_q)\}(y,y)}
                               {\tr\{ [(D_c + m_q)^{-1}]^{\dagger} (D_c + m_q)^{-1} \} (y,y)},   
\eea
where $ D^{-1}(m_q) $ is the sea quark propagator, and $ (D_c + m_q)^{-1} $ is the valence quark propagator.
Therefore, (\ref{eq:mres_delta}) is well-defined only in the unitary limit, with the valence quark mass
equal to the sea quark mass. We note that Eq. (\ref{eq:mres_delta}) is consistent with the form used in 
Refs. \cite{Brower:2004xi} and \cite{Edwards:2005an}, but not in the unitary limit.  
 
Similarly, the global residual mass (\ref{eq:def_Mres}) can be written as 
\bea
\label{eq:Mres_delta}
M_{res} = \frac{1}{4r} \frac{\Tr\{ [D^{-1}(m_q)]^{\dagger}(1-S^2) D^{-1}(m_q)\}}
                            {\Tr\{ [(D_c + m_q)^{-1}]^{\dagger} (D_c + m_q)^{-1} \}},  
\eea
where Tr denotes the trace over the Dirac, color, and site indices.

Nevertheless, it is tedious to compute the residual mass via (\ref{eq:mres_delta}) since it involves 
the multiplication of $ S = (1-\prod_s T_s)(1+ \prod_s T_s )^{-1} = H \sum_{i=1}^n b_i ( H^2 + d_i)^{-1} $ to 
the column vectors of $ D^{-1}(m_q) $, requiring conjugate gradient with multi-shift. 

In the following, we derive a practical formula for the residual mass, which only involves the valence quark propagator.   
Then the residual mass can be obtained once the valence quark propagator has been computed.
  
We observe that the numerator of (\ref{eq:mres_delta}) can be decomposed into two parts
\bea
\label{eq:Num}
  \tr \left\{ [D^{-1}(m_q)]^\dagger D^{-1}(m_q) \right\} (y,y)
- \tr \left\{ [S D^{-1}(m_q)]^\dagger (S D^{-1}(m_q)) \right\} (y,y). 
\eea
Using (\ref{eq:DcmI}) and (\ref{eq:Dc}), we obtain 
\BAN
S = \gamma_5 \left[ 2 r \frac{ D(m_q) - m_q }{ 1 - r m_q } - 1 \right],  
\EAN 
and 
\BAN
\label{eq:SD}
S D(m_q)^{-1} 
&=& \gamma_5 \left[ \frac{2r}{1 - r m_q} - \frac{ 1 + r m_q }{ 1 - r m_q } D(m_q)^{-1} \right] \nn
&=& \gamma_5 \left[ r - \left( 1 + r m_q \right) (D_c+m_q)^{-1} \right]. 
\EAN
Thus the second term in (\ref{eq:Num}) can be evaluated as  
\bea
\label{eq:Num2}
&& \tr \left\{ [S D^{-1}(m_q))]^\dagger 
                        (S D^{-1}(m_q) \right\}(y,y)  \nn
&& = r^2 \tr\Id - 2r(1+rm_q) \re \ \tr(D_c+m_q)^{-1}(y,y)  \nn
&& \hspace{10mm}  + (1+r m_q)^2 \tr\{ [(D_c + m_q)^{-1}]^\dagger (D_c + m_q)^{-1} \}(y,y). 
\eea
Using (\ref{eq:DcmI}), the first term of (\ref{eq:Num}) is evaluated as  
\bea
\label{eq:Num1}
&&   \tr \left\{ [D^{-1}(m_q)]^\dagger D^{-1}(m_q) \right\}(y,y) \nn
&& = r^2 \tr\Id + 2r(1-rm_q) \re \ \tr(D_c+m_q)^{-1}(y,y) \nn  
&& \hspace{10mm}  + (1-r m_q)^2 \tr\left\{ [(D_c + m_q)^{-1}]^\dagger (D_c + m_q)^{-1} \right\}(y,y).  
\eea
Substituting (\ref{eq:Num1}) and (\ref{eq:Num2}) into (\ref{eq:mres_delta}), we obtain
a formula for the residual mass 
\bea
\label{eq:mres_Dc}
m_{res}(y) 
= \frac{ \re \ \tr\{(D_c + m_q)^{-1}(y,y)\} }{ \tr\{ [(D_c + m_q)^\dagger (D_c+m_q)]^{-1}(y,y)\} } - m_q, 
\eea
which immediately gives the residual mass once the 12 columns 
of the valence quark propagator $ (D_c + m_q)^{-1}(x,y) $ have been computed. 
Also, it is appealing from the viewpoint of exact chiral symmetry, 
since the first term in (\ref{eq:mres_Dc}) gives $ m_q $ when $ D_c $ is exactly chirally symmetric, 
(i.e. $ D_c \gamma_5 + \gamma_5 D_c = 0 $), thus the residual mass is exactly zero. 

Similarly, the global residual mass (\ref{eq:Mres_delta}) can be written as
\bea
\label{eq:Mres_Dc}
M_{res} 
= \frac{ \re \ \Tr\{(D_c + m_q)^{-1}\} }{ \Tr\{ [(D_c + m_q)^\dagger (D_c+m_q)]^{-1}\} } - m_q. 
\eea
Equations (\ref{eq:mres_Dc}) and (\ref{eq:Mres_Dc}) are two of the main results of this paper.

Now we consider an ensemble of gauge configurations generated in full QCD with $ n_f $ flavors, 
obeying the probability distribution 
\BAN
\prod_{f=1}^{n_f} \det D(m_f) e^{-{\cal A}_g }, 
\EAN
then the ensemble average of the residual mass can be written as  
\BAN
\langle m_{res}(y) \rangle = 
\frac{ \int [dU]  \prod_{f=1}^{n_f} \det D(m_f) e^{-{\cal A}_g} m_{res}(y) } 
     { \int [dU]  \prod_{f=1}^{n_f} \det D(m_f) e^{-{\cal A}_g} },  
\EAN
which would be independent of $ y $ if the number of gauge configurations is sufficiently large.
Obviously, the ensemble average of the global residual mass, $ \langle M_{res} \rangle $, 
would tend to the limiting value with much less number of configurations.

\section{An upper bound for the residual mass}
\label{upper_bound_mres}

For ODWF, $ S(H) = S_{opt}(H) $, the Zolotarev optimal rational approximation
of $ \mbox{sgn}(H)= H(H^2)^{-1/2} $, provided that the eigenvalues of $ H^2 $ lying in 
the range $ [ \lambda_{min}^2, \lambda_{max}^2 ] $, where $ \lambda_{min}^2 $ and $ \lambda_{max}^2 $
are the lower and upper bounds used for computing the nonzero weights 
$ \{ \omega_s, s =1, \cdots, n-1, n+2, N_s+2 \} $.
Thus, for any gauge configuration yielding eigenvalues of $ H^2 $ lying in the range
$ [ \lambda_{min}^2, \lambda_{max}^2 ] $, the residual mass must be bounded since it is a function
of the sign function error $ || 1 - S_{opt}(H) || $ which is always less than   
$ d_Z $, the maximum deviation in the Zolotarev optimal rational approximation.
In the following, we obtain an upper bound for the global residual mass in lattice QCD with ODWF.

The numerator of (\ref{eq:Mres_delta}) can be written as  
\bea
\label{eq:inequality}
\Tr\{[D(m_q)^{-1}]^\dagger (1-S^2) D^{-1}(m_q)\} 
&=& \Tr\{(1-S^2) {(D^\dagger D)}^{-1}(m_q)\}  \nn
&\le& \left| \Tr\{ (1-S^2) {(D^\dagger D)}^{-1}(m_q)\} \right| \le \sum_j \alpha_j \beta_j,   
\eea 
where the von Neumann's trace inequality has been used, and 
$ \alpha_j $ and $ \beta_j $ are the eigenvalues of $ |1-S^2| $ and $ (D^\dagger D)^{-1}(m_q) $ respectively, 
in the ascending order, i.e., $ \alpha_1 \le \alpha_2 \le \cdots \le \alpha_N $, and 
$ \beta_1 \le \beta_2 \le \cdots \le \beta_N $.

For ODWF, $ S=S_{opt} $, $ \alpha_j \le 2 d_Z $. Thus (\ref{eq:inequality}) gives 
\bea
\Tr\{[D(m_q)^{-1}]^\dagger (1-S_{opt}^2) D^{-1}(m_q)\} \le 2 d_Z \Tr\{{(D^\dagger D)}^{-1}(m_q)\},   
\eea 
and (\ref{eq:Mres_delta}) implies 
\bea
\label{eq:Mres_opt}
M_{res} \le \frac{d_Z}{2r} \frac{\Tr\{[D(m_q)^{-1}]^\dagger D^{-1}(m_q)\}}
                                {\Tr\{[(D_c + m_q)^{-1}]^\dagger (D_c + m_q)^{-1}\}}. 
\eea

From (\ref{eq:DcmI}), the singular values of $ D^{-1}(m_q) $ and $ (D_c + m_q)^{-1} $ have the following relationship
\bea
\label{eq:lambda_xi}
\lambda_j = r + (1-r m_q) \xi_j, 
\eea 
where $ \xi_j $ is a singular value of $ (D_c + m_q)^{-1} $ and $ \lambda_j $ is the corresponding singular value of $ D^{-1}(m_q) $.
Then (\ref{eq:lambda_xi}) gives 
\bea
\frac{\sum_j |\lambda_j|^2}{\sum_j |\xi_j|^2} = (1-rm_q)^2 
                                                + 2 r(1-rm_q) \frac{ \langle \re (\xi) \rangle}{ \langle |\xi|^2 \rangle }
                                                + \frac{r^2}{\langle |\xi|^2 \rangle}, 
\eea
where 
\bea
\langle |\xi|^2 \rangle = \frac{1}{N} \sum_{j=1}^N |\xi_j|^2, \hspace{4mm} 
\langle \re (\xi) \rangle = \frac{1}{N} \sum_{j=1}^N \re (\xi_j), 
\eea  
and $ N $ is the total number of singular values of $ (D_c + m_q)^{-1} $. 
Therefore
\bea
\label{eq:ratio_bound}
\frac{\Tr\{[D(m_q)^{-1}]^\dagger D^{-1}(m_q)\}} {\Tr\{[(D_c + m_q)^{-1}]^\dagger (D_c + m_q)^{-1}\} }
%= \frac{ \sum_j | \lambda_j |^2 }{\sum_j |\xi_j|^2} 
= (1-rm_q)^2  + 2 r(1-rm_q) \frac{ \langle \re (\xi) \rangle}{ \langle |\xi|^2 \rangle }
                                                + \frac{r^2}{\langle |\xi|^2 \rangle},   
\eea
and (\ref{eq:Mres_opt}) becomes 
\bea
\label{eq:Mres_bound}
M_{res} \le \frac{d_Z}{2r} \left[ (1-rm_q)^2 + 2 r(1-rm_q) \frac{ \langle \re (\xi) \rangle}{ \langle |\xi|^2 \rangle }
                                                + \frac{r^2}{\langle |\xi|^2 \rangle} \right]. 
\eea 
%for any gauge configuration yielding the eigenvalues of $ H^2 $ lying within the range 
%$ [ \lambda_{min}^2, \lambda_{max}^2 ] $, where $ \lambda_{min}^2 $ and $ \lambda_{max}^2 $ are 
%the upper-bound and lower-bound used for computing the optimal weights $ \{ \omega_s \} $. 
Thus, to obtain the upper bound of $ M_{res} $ amounts to 
working out an upper bound for $ \langle \re (\xi) \rangle/\langle |\xi|^2 \rangle $, 
and a lower bound for $ \langle |\xi|^2 \rangle$. 

First we work out a lower bound for $ \langle |\xi|^2 \rangle$. 
The eigenvalues of $ V = \gamma_5 S $ can be expressed as $\{ R_j e^{i\theta_j}, j=1,\cdots, N \} $, 
where $ R_j $ can be bigger or less than one since the chiral symmetry is not exact for finite $ N_s $.
Then the corresponding eigenvalues of $ r D_c = (1+V)(1-V)^{-1} $ can be expressed as 
\bea
\label{eq:rDc_xy}
x_j+iy_j
%=\frac{1+R_j\cos\theta_j+iR_j\sin\theta_j}{1-R_j\cos\theta_j-iR_j\sin\theta_j}
=\frac{1-R_j^2+i2R_j\sin\theta_j}{1+R_j^2 - 2R_j\cos\theta_j}. 
%=\frac{1-R_j^2+i2R_j\sin\theta_j}{(1-R_j\cos\theta_j)^2+R_j^2(1-\cos^2\theta_j)}. 
\eea
%which implies that 
%\bea
%R_j > 1 \Leftrightarrow x_j < 0, \nonumber\\ %R_j = 1 \Leftrightarrow x_j = 0, \nonumber\\
%R_j < 1 \Leftrightarrow x_j > 0. \nonumber
%\eea
Thus the eigenvalues of $ (D_c + m_q)^{-1} $ are $ \eta_j = r ( x_j + r m_q + i y_j )^{-1} $.
For finite $ N_s $, $ \langle |\eta|^2 \rangle $ is not exactly equal to
$ \langle |\xi|^2 \rangle $, since $ [V^{\dagger}, V] \ne 0 $, due to the eigenvalues of $ V $
not falling on a circle.   
However, in estimating the lower bound of $ \langle |\xi|^2 \rangle $, one must fix
all eigenvalues of $ V $ on a circle with a radius having the maximal deviation from one.
Then, in this case, $ [ V^\dagger, V ] = 0 $, and    
$ \langle |\eta|^2 \rangle = \langle |\xi|^2 \rangle $. 
Thus, we can use $ \langle |\eta|^2 \rangle $ to estimate the lower bound of $ \langle |\xi|^2 \rangle $. 
Using (\ref{eq:rDc_xy}) and setting $ R_j = R $, and $ m \equiv r m_q $, we obtain
\bea
\langle |\xi|^2 \rangle &=& \frac{1}{N} \sum_{j=1}^N \frac{r^2}{(x_j + m )^2 + y_j^2}    
= \frac{r^2}{N} \sum_{j=1}^N \frac{(1+R^2-2R \cos\theta_j)} 
                                  {(1+m^2)(1+R^2) + 2m(1-R^2) + 2 R (1-m^2)\cos\theta_j}, \nn 
%&=& -\frac{r^2}{1-m^2} + \frac{2r^2}{1-m^2} \left[\frac{1+R^2 + m(1-R^2)}{(1+m^2)(1-R^2) + 2m(1+R^2)} \right],  \\
&=&  \frac{r^2}{1+m} \left[\frac{2(1+R^2)-(1-m)(1-R^2)}{(1+m^2)(1-R^2) + 2m(1+R^2)} \right], 
\label{eq:xi2}  
\eea  
where we have assumed that the distribution of the eigenvalues of $ V $ is uniform in $ \theta $, and used the formula
\BAN
\frac{1}{2\pi} \int_0^{2\pi} \frac{d\theta}{B+A\cos\theta} &=& \frac{1}{\sqrt{B^2 - A^2}}.   
%\frac{1}{2\pi} \int_0^{2\pi} \frac{d\theta \cos\theta}{B+A\cos\theta} &=& \frac{1}{A} \left(1 - \frac{B}{\sqrt{B^2 - A^2}} \right).
\EAN   
For ODWF, $ |1-R^2| \le 2 d_Z $, and the lower bound of (\ref{eq:xi2}) is attained at
$ R = \sqrt{1-2 d_Z } $, i.e., 
\bea
\label{eq:xi2_lower_bound}
\langle |\xi|^2 \rangle     
%\ge  \frac{r^2}{1-m^2} \left[\frac{4(1-m)(1-d_Z)-2(1-m)^2 d_Z}{2(1+m^2) d_Z + 4m(1-d_Z)} \right]  
\ge \frac{r^2}{1+m} \left[ \frac{2-(3-m)d_Z}{2m + (1-m)^2 d_Z} \right]. 
\eea

Next we work out an upper bound for $ \langle \re (\xi) \rangle/\langle |\xi|^2 \rangle $.
Again, using (\ref{eq:rDc_xy}) and setting $ R_j = R $, and $ m \equiv r m_q $, we obtain
\bea
\langle \re (\xi) \rangle   
&=& \frac{1}{N} \sum_{j=1}^N \frac{r(x_j+m)}{(x_j + m )^2 + y_j^2}    
= \frac{r}{N} \sum_{j=1}^N \frac{1-R^2+m(1+R^2-2R\cos\theta_j)} 
                                  {(1+m^2)(1+R^2) + 2m(1-R^2) + 2 R (1-m^2)\cos\theta_j}, \nn 
&=& r \frac{1-R^2}{(1+m^2)(1-R^2) + 2m(1+R^2)} + r^{-1} m \langle |\xi|^2 \rangle,   
\label{eq:Re(xi)}  
\eea
and
\bea
\label{eq:ratio}
\frac{ \langle \re (\xi) \rangle}{ \langle |\xi|^2 \rangle } = r^{-1} \left[ m + \frac{(1+m)(1-R^2)}{2(1+R^2)-(1-m)(1-R^2)} \right], 
\eea
where (\ref{eq:xi2}) has been used.

For ODWF, $ |1-R^2| \le 2 d_Z $, the upper bound of (\ref{eq:ratio}) is attained at
$ R = \sqrt{1-2 d_Z } $, i.e., 
\bea
\label{eq:ratio_upper_bound}
\frac{ \langle \re (\xi) \rangle}{ \langle |\xi|^2 \rangle } \le r^{-1} \left[ m + \frac{(1+m)d_Z}{2-(3-m)d_Z} \right].    
\eea

Substituting (\ref{eq:ratio_upper_bound}) and (\ref{eq:xi2_lower_bound}) into (\ref{eq:Mres_bound}), we obtain
\bea
\label{eq:Mres_upper_bound}
%\begin{aligned}
%M_{res} &\le \frac{d_Z}{2r} \left\{ (1-m)^2 + 2 (1-m) \left[ m + \frac{(1+ m)d_Z}{2-(3- m)d_Z} \right]  \right.  \\    
%        & \hspace{50mm}   \left.             + (1+m) \left[\frac{2m + (1-m)^2 d_Z}{2-(3-m) d_Z} \right] \right\}.
M_{res} \le \frac{d_Z}{2r} \left[ \frac{2(1+m)}{2-(3-m)d_Z} \right], 
%\end{aligned}
\eea
where $ m \equiv r m_q $.
This is one of the main results of this paper. 

In general, $ 0 \le d_Z \le 0.5 $, this gives 
\bea
\label{eq:factor_m}
(1+m) \le \frac{2(1+m)}{2-(3-m)d_Z} \le 4. 
\eea
Thus the upper-bound of $ M_{res} $ varies in the range
\bea
\frac{d_Z}{2r} (1+m) \le (M_{res})^{upper-bound} \le  \frac{2 d_Z}{r}.
\eea
  
For ODWF, $ d_Z \ll 1 $ for $ N_s \gg 1 $, 
then (\ref{eq:Mres_upper_bound}) reduces to    
\bea
M_{res} \le 
\frac{d_Z}{2r} \left( 1 + r m_q \right) \simeq  \frac{d_Z}{2r},   
\label{eq:Mres_upper_bound_A}
\eea
where $ m_q \ll m_{PV} = r^{-1} $ has been used in the last approximation.    

Moreover, $ d_Z $ is an exponentially decreasing function of $ N_s $, 
and it can be parametrized as \cite{Chiu:2002eh}
\bea
\label{eq:eZ}
d_Z(N_s,b) =  A(b) \exp\{ - C(b) N_s \}, \hspace{4mm} b \equiv \lambda_{max}^2/\lambda_{min}^2, 
%\ N_s = 2n, 
\eea
where $ A(b) $ and $ C(b) $ are positive definite functions of $ b $. 
%\BAN
%\label{eq:cb}
% C(b) &=& 4.58(5) [\ln (b)]^{-0.774(5)},  \\
%\label{eq:Ab}
% A(b) &=& 0.93(1) [\ln (b)]^{0.596(9)}, 
%\EAN
This immediately implies that the global residual mass for lattice QCD with ODWF is  
an exponentially decreasing function of $ N_s $, regardless of the size of the lattice, 
at zero or finite temperatures. However, this scenario holds only when all eigenvalues 
of $ H^2 $ are falling inside the interval $ [ \lambda_{min}^2, \lambda_{max}^2 ] $. 
In general, in the simulation of full QCD with ODWF, after fixing $ \lambda_{min}^2 $ 
and $ \lambda_{max}^2 $ in the beginning of the simulation, it could  
happen that some (accepted) gauge configurations in the course of the simulation may yield 
eigenvalues of $ H^2 $ lying outside the interval $ [\lambda_{min}^2, \lambda_{max}^2] $.
Then the global residual mass of such ``unbounded" configurations could be 
larger than the upper bound (\ref{eq:Mres_upper_bound}), especially for those
with many eigenvalues of $ H^2 $ less than $ \lambda_{min}^2 $. 
Thus, after generating an ensemble of gauge configurations, 
the ensemble averaged residual mass, $ \langle m_{res}(y) \rangle $ or
$ \langle M_{res} \rangle $, could be larger than the upper-bound (\ref{eq:Mres_upper_bound}). 
%if the number of "unbounded" eigenvalues of $ H^2 $  
%in many configurations is not sufficiently small.
In the following, we discuss to what extent the low-lying eigenmodes of $ H^2 $ modify the 
upper-bound (\ref{eq:Mres_upper_bound}).

Consider a configuration $U$ of which $ |H| $ has $ N_a $ eigenvalues ($ h_i, i=1,\cdots,N_a $) 
less than $ \lambda_{min} $, i.e., $ h_1 < h_2 < \cdots < h_{N_a} < \lambda_{min} $. 
Here we assume $ N_a \ll N = 12 L^3 T $.   
For each of these $ N_a $ eigenvalues, the corresponding eigenvalue of $ |1 - S_{opt}^2(H)| $  
is greater than $ 2 d_Z $, with the maximum equal to  
\bea
\label{eq:da}
2 d_a \equiv | 1 - S_{opt}^2(h_1) |,   
\eea
where $ h_1 $ is the smallest eigenvalue of $ |H| $. 
Therefore (\ref{eq:inequality}) is modified to 
\bea
\Tr\{[D(m_q)^{-1}]^\dagger (1-S_{opt}^2) D^{-1}(m_q)\}
&\le& 2d_Z \sum_{j=1}^{N-N_a} \beta_j + 2 d_a \sum_{j=N-N_a+1}^{N} \beta_j \nonumber\\
&=& [ 2(d_a - d_Z) Q_a + 2 d_Z ] \Tr\{{(D^\dagger D)}^{-1}(m_q)\},
\eea
where 
\bea
Q_a &\equiv& \frac{\sum_{j=N-N_a+1}^{N} \beta_j}{\sum_{j=1}^{N} \beta_j}, \\
\sum_{j=1}^{N} \beta_j &=& \Tr\{{(D^\dagger D)}^{-1}(m_q)\}. 
\eea
Now the upper-bound of $ \Tr \{ {(D^\dagger D)}^{-1}(m_q) \} / \Tr\{[(D_c + m_q)^{-1}]^\dagger (D_c + m_q)^{-1}\} $
can be evaluated as before, except replacing $ d_Z $ with $ d_a $. 
Then the upper-bound of the global residual mass (\ref{eq:Mres_upper_bound}) is transcribed to  
\bea
\label{eq:Mres_upper_bound_Q}
%\begin{aligned}
%M_{res} &\le \left[ \frac{(d_a - d_Z) Q_a + d_Z }{2r} \right] 
%             \left\{ (1-m)^2 + 2 (1-m) \left[ m + \frac{(1+ m)d_a}{2-(3- m)d_a} \right]  \right.  \\    
%   & \hspace{50mm}   \left.    + (1+m) \left[\frac{2m + (1-m)^2 d_a}{2-(3-m) d_a} \right] \right\} \equiv F(N_s,m,N_a,h_1).
M_{res} \le \left[ \frac{d_Z+(d_a - d_Z) Q_a }{2r} \right] \left[ \frac{2(1+m)}{2-(3-m)d_a} \right] \equiv F(N_s, m, N_a, h),   
%\end{aligned}
\eea
where the factor $ 2(1+m)/[2-(3-m)d_a] $ is bounded between $ (1+m) $ and 4, similar to (\ref{eq:factor_m}).
Thus the most significant change due to the ``unbounded" low-lying eigenmodes is to replace 
$ d_Z $ with $ d_Z + (d_a-d_Z)Q_a $, in the first factor of (\ref{eq:Mres_upper_bound_Q}).
  
Next we evaluate $Q_a$. Using (\ref{eq:lambda_xi}), we obtain
\bea
\label{eq:Qa}
Q_a
=\frac{\sum_{j=N-N_a+1}^{N} \beta_j}{\sum_{j=1}^{N} \beta_j} 
= \frac{(N_a/N) r^2 + 2r(1- m)\lceil \re(\xi) \rfloor + (1-m)^2 \lceil |\xi|^2 \rfloor}
       {r^2 + 2r(1- m)\langle \re (\xi) \rangle + (1-m)^2 \langle |\xi|^2 \rangle}, 
%\hspace{4mm} f_a \equiv \frac{N_a}{N}
\eea
where $ \xi_j $ is a singular value of $ (D_c + m_q)^{-1} $, and 
\bea
&& \lceil |\xi|^2 \rfloor = \frac{1}{N} \sum_{j=N-N_a+1}^{N} |\xi_j|^2, \hspace{4mm}
\lceil \re (\xi) \rfloor = \frac{1}{N} \sum_{j=N-N_a+1}^{N} \re (\xi_j),\\
&& \langle |\xi|^2 \rangle = \frac{1}{N} \sum_{j=1}^N |\xi_j|^2, \hspace{4mm}
\langle \re (\xi) \rangle = \frac{1}{N} \sum_{j=1}^N \re (\xi_j). 
\eea
To estimate above sums, we follow the same procedure in obtaining (\ref{eq:xi2})   
and (\ref{eq:Re(xi)}), and also assume that the distribution of the eigenvalues 
of $ V = \gamma_5 S_{opt} $ is uniform in $ \theta $. Then we have 
\bea
\langle |\xi|^2 \rangle    
&=&  \frac{r^2}{1+m} \left[\frac{2(1+R^2)-(1-m)(1-R^2)}{(1+m^2)(1-R^2) + 2m(1+R^2)} \right], \\ 
\langle \re (\xi) \rangle   
&=& r \frac{1-R^2}{(1+m^2)(1-R^2) + 2m(1+R^2)} + r^{-1} m \langle |\xi|^2 \rangle,   \\   
\label{eq:xi2_A}
\lceil |\xi|^2 \rfloor 
&=& \frac{r^2}{\pi} \int_{\theta_a}^{\pi} d\theta \frac{1+R^2-2R \cos\theta}
                                     {(1+m^2)(1+R^2) + 2m(1-R^2) + 2 R (1-m^2)\cos\theta},\\
\label{eq:Rexi_A}
\lceil \re (\xi) \rfloor
&=& \frac{r}{\pi} \int_{\theta_a}^{\pi} d\theta \frac{1-R^2+m(1+R^2-2R\cos\theta)}
                                   {(1+m^2)(1+R^2) + 2m(1-R^2) + 2 R (1-m^2)\cos\theta},
\eea
where $ R = \sqrt{1-2 d_a} $, and $ \theta_a = ( 1 - N_a/N) \pi $.
Then the denominator of $ Q_a $ becomes 
\bea
\label{eq:Qa_denominator}
r^2 + 2r(1- m)\langle \re (\xi) \rangle + (1-m)^2 \langle |\xi|^2 \rangle = \frac{2 r^2}{2m + (1-m)^2 d_a}.
\eea
To evaluate the integrals (\ref{eq:xi2_A}) and (\ref{eq:Rexi_A}), we perform the change of variable $ \chi = \pi - \theta $, 
and obtain
\bea
\label{eq:xi2_A1}
\lceil |\xi|^2 \rfloor 
&=& \frac{r^2}{\pi} \int_{0}^{N_a \pi/N} d\chi \frac{1+R^2-2R \cos(\pi-\chi)}
                                     {(1+m^2)(1+R^2) + 2m(1-R^2) + 2 R (1-m^2)\cos(\pi-\chi)},\\
\label{eq:Rexi_A1}
\lceil \re (\xi) \rfloor
&=& \frac{r}{\pi} \int_{0}^{N_a \pi/N} d\chi \frac{1-R^2+m(1+R^2-2R\cos(\pi-\chi))}
                                   {(1+m^2)(1+R^2) + 2m(1-R^2) + 2 R (1-m^2)\cos(\pi-\chi)}.
\eea
Since the upper limit of the integrals is $ N_a \pi /N \ll 1 $, we can use the approximation \\
$ \cos(\pi - \chi) \simeq -1 + \chi^2/2 $ in the integrand, and obtain
\bea
&&\lceil |\xi|^2 \rfloor = \frac{N_a}{N} \left( \frac{r d_a}{m d_a+(1-d_a-\sqrt{1-2d_a})}\right)^2, \\
\nn
&&\lceil \re (\xi) \rfloor= \frac{N_a}{N} \frac{r d_a}{m d_a+(1-d_a-\sqrt{1-2d_a})}, 
\eea   
where we have used the formula
\BAN
\int_0^{\chi_a} d \chi \frac{A+B \chi^2}{C+D \chi^2} 
= \frac{B}{D} \chi_a - \frac{(BC-AD) \tan^{-1} \left( \sqrt{D/C} \chi_a \right)}{D\sqrt{CD}} 
%&\approx& \frac{B}{D} \chi_a - \frac{(BC-AD) \sqrt{D/C} \chi_a }{D\sqrt{CD}}\nn
\simeq \frac{A}{C} \chi_a + {\cal O}(\chi_a^2), 
\EAN
and suppressed the higher order terms of $ {\cal O}( (N_a/N)^2 ) $. 
Then we obtain the numerator of $ Q_a $ 
\bea
\label{eq:Qa_numerator}
&&  \left(\frac{N_a}{N}\right)r^2 + 2r(1-m)\lceil \re(\xi) \rfloor +(1-m)^2 \lceil |\xi|^2 \rfloor \nn 
&=& \left(\frac{N_a}{N}\right)r^2 \left( \frac{1-\sqrt{1-2d_a}}{1-\sqrt{1-2d_a}-(1-m)d_a} \right)^2. 
\eea 
Using (\ref{eq:Qa_denominator}) and (\ref{eq:Qa_numerator}), we get 
\bea
\label{eq:Qa_final}
Q_a 
&=& \frac{N_a}{N}\left( m+(1-m)^2\frac{d_a}{2} \right) \left( \frac{1-\sqrt{1-2d_a}}{1-\sqrt{1-2d_a}-(1-m)d_a} \right)^2 \nn
%&=& \frac{N_a}{N}\left( 2m+(1-m)^2d_a \right) \frac{1-\sqrt{1-2d_a}-d_a}{\left[1-\sqrt{1-2d_a}-(1-m)d_a\right]^2} \nn
&=& \frac{N_a}{N}\left[ \frac{(1+m)+(1-m)\sqrt{1-2d_a}}{(1+m)-(1-m)\sqrt{1-2d_a}} \right].
\eea

Since $ 0 \le d_a \le 0.5 $, we have  
\bea
\frac{N_a}{N} \le Q_a \le \frac{N_a}{N} \left( \frac{1}{m} \right). 
\eea
In other words, $ Q_a $ is a monotonically decreasing function of $ d_a $, which in turn is a monotonically increasing function 
of $ N_s $, with the upper bound $ (N_a/N)/m $. 

For $ N_s \gg 1 $, $ d_Z \ll d_a \ll 0.5 $, then 
\bea
\label{eq:daQa}
(d_a - d_Z) Q_a \simeq d_a Q_a \simeq \frac{N_a}{N} \left( \frac{d_a}{m + d_a/2} \right), 
\eea
where $ m \ll 1 $ has been used. If $ m \ll d_a/2 $, then (\ref{eq:daQa}) gives $ (d_a - d_Z)Q_a \simeq 2 N_a/N $, which 
is almost independent of $ N_s $. This immediately implies that the first factor \\ 
$ [d_Z + (d_a - d_Z) Q_a]/(2r) $ 
in the upper-bound of the residual mass (\ref{eq:Mres_upper_bound_Q}) would look like almost 
``saturated" (decreasing slowly with respect to $ N_s $) 
after $ N_s $ greater than some threshold value $ N_s^{thres} $ 
which depends on $ N_a $ (the number of eigenvalues of $ |H| $ smaller than $ \lambda_{min} $) 
and weakly on $ h_1 $ (the smallest eigenvalue of $ |H| $). 
In other words, if there are some eigenvalues of $ |H| $ smaller than $ \lambda_{min} $,  
the exponential bound cannot be sustained for $ N_s > N_s^{thres} $. 
This is one of the most interesting results emerging from our theoretical analysis.
 
\begin{figure*}[tb]
\begin{center}
%\begin{tabular}{@{}c@{}c@{}}
\includegraphics*[width=10cm,clip=true]{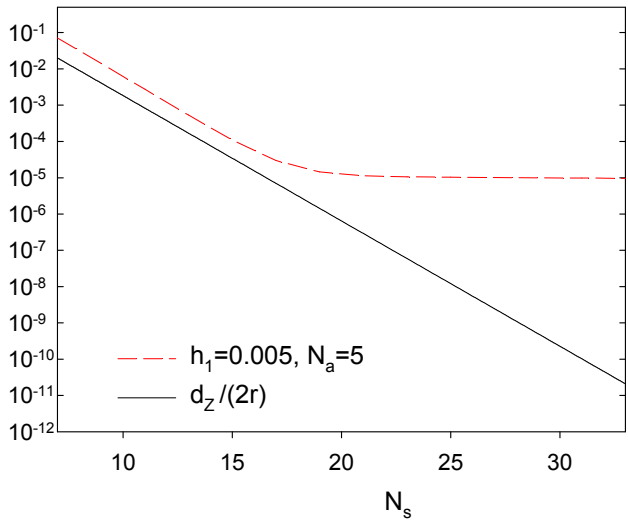}
%&
%\includegraphics*[width=7.5cm,clip=true]{Fpi_mq_nf2_b595_GeV.eps}
%\\ (a) & (b)
%\end{tabular}
\caption{The solid line is the theoretical upper-bound (\ref{eq:Mres_upper_bound_A}) of the global residual mass, 
         for the case when all eigenvalues of $ |H| $ fall inside the interval $ [\lambda_{min}, \lambda_{max}] $.   
         Here we set $ \lambda_{min} = 0.05 $, and $ \lambda_{max} = 6.20 $.
         The dotted line is the modified upper-bound (\ref{eq:Mres_upper_bound_Q}) for the 
         case when some of the eigenvalues of $ |H| $ are smaller than $ \lambda_{min} $.  
         Here we set $ N_a=5 $ (i.e., $ 5 $ eigenvalues of $ |H|$ smaller than $ \lambda_{min} $), 
         and the smallest eigenvalue $ h_1 = 0.005 $.}
\label{fig:Mres_Qa_Ns}
\end{center}
\end{figure*}

In Fig. \ref{fig:Mres_Qa_Ns}, we plot the upper-bounds
(\ref{eq:Mres_upper_bound_A}) and (\ref{eq:Mres_upper_bound_Q}) versus $ N_s $ respectively, 
where the integrals in $ Q_a $ are evaluated numerically. 
Here we set $ \lambda_{min} = 0.05 $, and $ \lambda_{max} = 6.20 $.
The solid line is the upper-bound (\ref{eq:Mres_upper_bound_A}) of the global residual mass, 
for the case when all eigenvalues of $ |H| $ fall inside the interval 
$ [\lambda_{min}, \lambda_{max}] $. It decays exponentially with $ N_s $.   
The dotted line is the modified upper-bound (\ref{eq:Mres_upper_bound_Q})
of the global residual mass, for the case when 
some eigenvalues of $ |H| $ are smaller than $ \lambda_{min} $. 
Here we set $ N_a=5 $ (i.e., $ 5 $ eigenvalues of $ |H|$
smaller than $ \lambda_{min} $), and the smallest eigenvalue 
$ h_1 = 0.005 = \lambda_{min}/10 $.
We see that the modified upper-bound of the global residual mass of ODWF 
decays exponentially up to $ N_s \simeq 18 $, then it decays like
$ 1/N_s^{3} $ for $ N_s \simeq 18-23 $, and almost ``saturates" at  
$ \sim 10^{-5} $ for $ N_s \simeq 24-33 $.  
This agress with our theoretical analysis of the first factor $ [d_Z + (d_a - d_Z) Q_a]/(2r) $ 
in the upper-bound of the residual mass (\ref{eq:Mres_upper_bound_Q}), which would become almost 
``saturated" when $ N_s > N_s^{thres} $.    
Moreover, due to the second factor $ 2(1+m)/[2-(3-m)d_a] $ in (\ref{eq:Mres_upper_bound_Q}), 
the exponential bound $ d_Z/(2r) $ for $ N_s < N_s^{thres} $ is increased by a factor $ \sim 3 $,   
as shown in Fig. \ref{fig:Mres_Qa_Ns}.
%We note that, for fixed $ N_s $, choosing a smaller $ \lambda_{min} $ 
%does not necessarily give a smaller global residual mass, since 
%$ d_Z/(2r) $ becomes larger for larger $ \lambda_{max}/\lambda_{min} $.

Next, we turn to an ensemble of gauge configurations $\{ U_i \}$.
Let the smallest eigenvalue of $ |H| $ with the gauge configuration $ U_i $ 
be $ h_1^{(i)} $, and the probability distribution of $ \{ h_1^{(i)} \} $ satisfies
\BAN
\int_{h_{min}}^{h_{max}} dh \rho(h) = 1, 
\EAN
where $ h_{min} = \min( h_1^{(1)}, h_1^{(2)}, \cdots, h_1^{(i)}, \cdots) $ 
and $ h_{max} = \max( h_1^{(1)}, h_1^{(2)}, \cdots, h_1^{(i)}, \cdots) $. 
Then the upper-bound of the global residual mass for an ensemble of gauge configurations is 
\bea
\label{eq:Mres_upper_bound_G}
\langle M_{res} \rangle \le \frac{1}{h_{max} - h_{min}} \left\{ \theta(x) \int_{h_{min}}^{y} dh \rho(h) F(N_s,m,N_a,h) 
                             + \left(\frac{d_Z}{2r}\right) \left[ z \theta(z) + x \theta(-x) \right] \right\}, 
\eea
where $ x = \lambda_{min} - h_{min} $, $ y = \min( \lambda_{min}, h_{max} ) $, $ z = h_{max} - \lambda_{min} $, 
and $ F(N_s, m, N_a, h) $ is defined in (\ref{eq:Mres_upper_bound_Q}).  
This is one of the main results of this paper.

\begin{figure*}[tb]
\begin{center}
%\begin{tabular}{@{}c@{}c@{}}
\includegraphics*[width=10cm,clip=true]{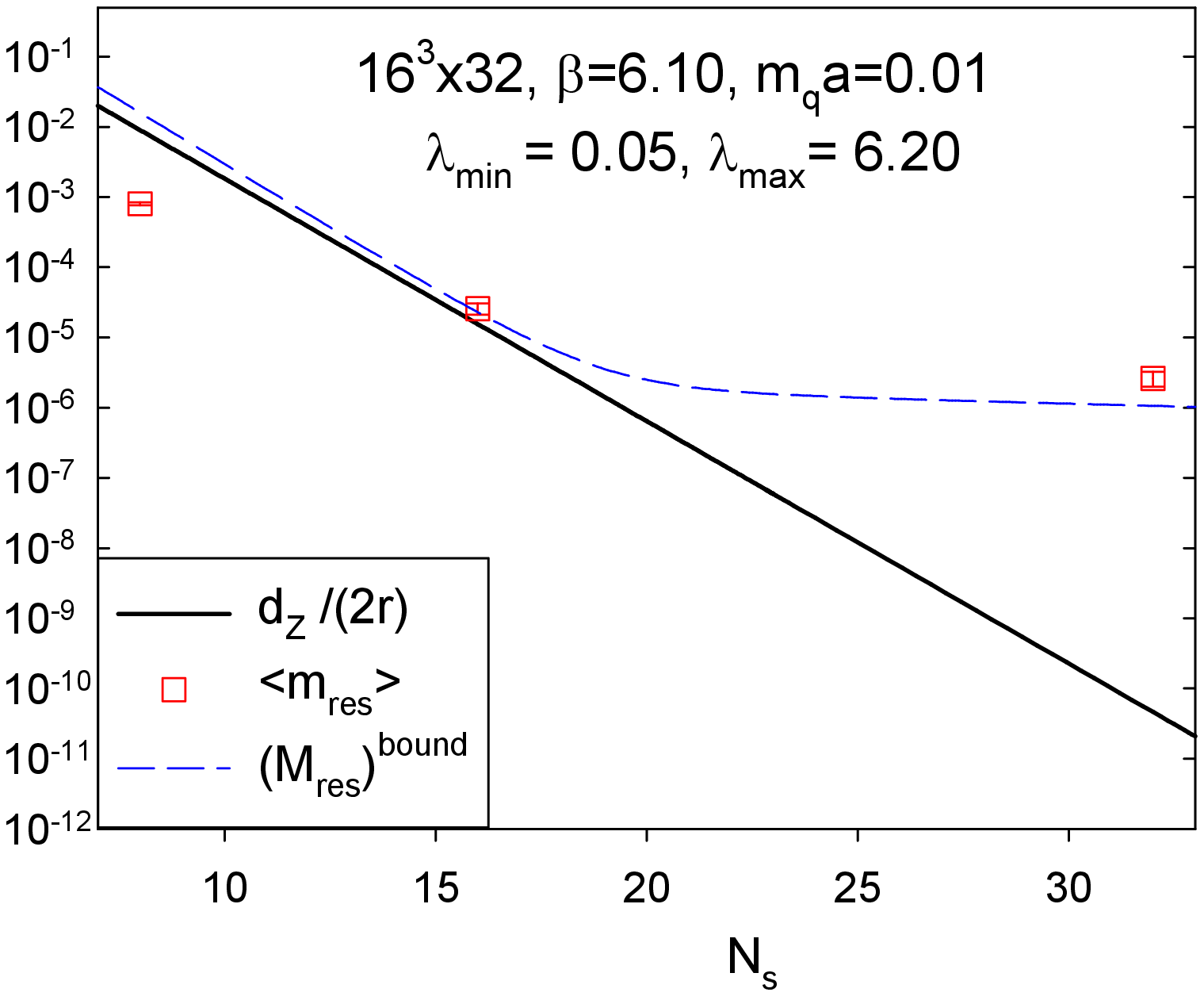}
%&
%\includegraphics*[width=7.5cm,clip=true]{Fpi_mq_nf2_b595_GeV.eps}
%\\ (a) & (b)
%\end{tabular}
\caption{The solid line is the theoretical upper-bound $ d_Z/(2r) $ of the global residual mass, for  
         the case when all eigenvalues of $ |H| $ fall inside the interval $ [\lambda_{min}, \lambda_{max}] $.   
         Here we set $ \lambda_{min} = 0.05 $, and $ \lambda_{max} = 6.20 $.
         The dotted line is the general upper-bound (\ref{eq:Mres_upper_bound_G}).
         The squares denote the average residual mass obtained with 243 gauge configurations from simulation of 
         2-flavors QCD with ODWF at $ \beta = 6.10 $, on the $ 16^3 \times 32 $ lattice.}
\label{fig:mres_b610_l16t32s16_m001_lmin005_lmax620}
\end{center}
\end{figure*}

\section{Numerical Tests}

In this section, we test to what extent the upper-bound (\ref{eq:Mres_upper_bound_G}) of global residual mass
for an ensemble of gauge configurations is satisfied in large-scale simulations of lattice QCD with ODWF. 
To this end, we perform hybrid Monte Carlo (HMC) 
simulations of two flavors QCD on the $ 16^3 \times 32 $ lattice with ODWF (setting kernel $ H = H_w $) 
at $ N_s = 16 $ and $ \lambda_{min}/\lambda_{max} = 0.05/6.20 $, plaquette gauge action at $ \beta = 6.10 $, 
and sea-quark mass $ m_q a = 0.01 $.  We have generated 2730 trajectories. 
After discarding the initial 300 trajectories for thermalization, 
we sample one configuration every 10 trajectories. Thus we have 243 gauge configurations.  
For each configuration, we compute the valence quark propagator with mass $ m_{val} a = m_{sea} a = 0.01 $, 
and use the formula (\ref{eq:mres_Dc}) to obtain the residual mass.
We perform the same calculation for 3 different cases: $ N_s = 8, 16, 32 $, with the same     
$ \lambda_{min}/\lambda_{max} = 0.05/6.20 $. Then we obtain the averaged residual mass $ \langle m_{res} \rangle $
for $ N_s = 8, 16, 32 $, which are denoted by squares in Fig. \ref{fig:mres_b610_l16t32s16_m001_lmin005_lmax620}.
We see that for $ N_s = 8 $ and $ N_s = 16 $, $ \langle m_{res} \rangle $ satisfies the 
exponential bound (\ref{eq:Mres_upper_bound_A}), $ d_Z/(2r) $. However, for $ N_s = 32 $, 
$ \langle m_{res} \rangle $ is much larger than the exponential bound.
 
\begin{figure*}[tb]
\begin{center}
%\begin{tabular}{@{}c@{}c@{}}
\includegraphics*[width=10cm,clip=true]{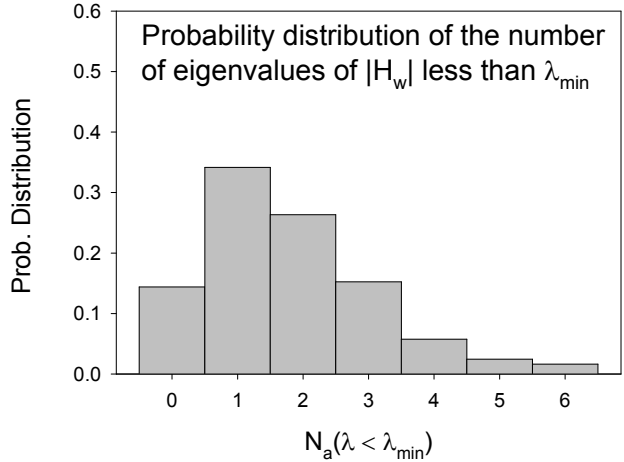}
%&
%\includegraphics*[width=7.5cm,clip=true]{distribution_Na.eps}
%\\ (a) & (b)
%\end{tabular}
\caption{The histogram of $ N_a $, the number of eigenvalues of $ |H_w| $ smaller than $ \lambda_{min} $, 
         for 243 gauge configurations generated by HMC simulation of two flavors QCD on the 
         $ 16^3 \times 32 $ lattice with ODWF at $ N_s = 16 $ and $ \lambda_{min}/\lambda_{max} = 0.05/6.20 $, 
         plaquette gauge action at $ \beta = 6.10 $, and sea-quark mass $ m_q a = 0.01 $.}
\label{fig:distribution_Na}
\end{center}
\end{figure*}

Next, we compute the general upper-bound (\ref{eq:Mres_upper_bound_G}) to see to what extent
it is satisfied by $ \langle m_{res} \rangle $. 
For each configuration, we project 250 lowest lying eigenmodes of $ |H_w| $.  
%using the adaptive thick-restart Lanczos algorithm \cite{a-TRLan}. 
Then we obtain the smallest eigenvalue ($ h_1 $) of $ |H_w| $, 
and also $ N_a $, the number of eigenmodes of $ |H_w| $ with eigenvalue smaller than $ \lambda_{min} $.
Among the set of 243 smallest eigenvalues $ \{ h_1^{(i)}, i=1,\cdots,243 \} $, 
the minimum is $ h_{min} = 6.99106 \times 10^{-5} <  \lambda_{min} $, 
while the maximum is $ h_{max} = 0.1186 > \lambda_{min} $.
The probability distribution of $ h_1 $ can be fitted by the ``log-normal" function
\bea
\label{eq:rho_b610}
\rho(h) =  \rho_0 \exp \left\{ -\frac{1}{2} \left[ \frac{\ln(h/h_0)}{\sigma} \right]^2 \right\}, 
\eea
where $ \rho_0 = 27.4326(1.2364) $, $ \sigma = 0.923(43) $, and $ h_0 = 0.0108(5) $.
The average number of eigenvalues of $ |H_w| $ smaller than $ \lambda_{min} $ is 
$ \langle N_a \rangle = 1.778(85) $. The histogram of $ N_a $ is plotted in 
Fig. \ref{fig:distribution_Na}. Using (\ref{eq:rho_b610}) and the information
of $ N_a $, we obtain the upper-bound (\ref{eq:Mres_upper_bound_G}) of the 
global residual mass as a function of $ N_s $, which is plotted as the 
dotted lines in Fig. \ref{fig:mres_b610_l16t32s16_m001_lmin005_lmax620}.
We see that the data points of $ \langle m_{res} \rangle $ are in good agreement 
with the upper-bound (\ref{eq:Mres_upper_bound_G}).

A salient feature emerging from this numerical study is that 
the exponential bound (\ref{eq:Mres_upper_bound_A}) for the global residual mass 
holds for $ N_s < 18 $.  
However, it is difficult to sustain the exponential bound for $ N_s > 20 $, 
due to the low-lying eigenmodes of $ |H| $ with eigenvalue less than $ \lambda_{min} $, 
in agreement with our theoretical analysis of the first factor $ [d_Z + (d_a - d_Z) Q_a]/(2r) $ 
in the upper-bound of the residual mass (\ref{eq:Mres_upper_bound_Q}), 
which would become almost ``saturated" when $ N_s > N_s^{thres} \simeq 18-20 $.    

At this point, it is instructive to compare the behavior of the residual 
mass of the ODWF in Fig. \ref{fig:mres_b610_l16t32s16_m001_lmin005_lmax620} 
with those of the conventional DWF.
For the conventional DWF, $ \langle m_{res} \rangle $
behaves like $ 1/N_s $ for the Shamir kernel \cite{Antonio:2008zz}, 
and $ 1/N_s^2 $ for the M\"obius kernel with tuned parameters \cite{Brower:2012vk}.
However, they do not possess an exponential bound for any interval of $ N_s $, 
unlike the case of ODWF. Moreover, it is straightforward to generalize the theoretical analysis 
in the last section to the case of conventional DWF and show that the ``saturation" phenomenon
at large $ N_s $ also holds for the conventional DWF.

\section{Concluding remarks}

In this paper, we have derived the axial Ward identity for lattice QCD with ODWF, 
by introducing two transparent layers at the central region of the fifth dimension, 
in addition to the two transparent layers at the boundaries 
for defining the quark fields \cite{Chiu:2003ir}. 
From the axial Ward identity (\ref{eq:AWI_O_Nf}), we obtain (\ref{eq:def_mres_O}) and 
(\ref{eq:def_Mres_O}) as the local and global residual mass, 
for measuring the chiral symmetry breaking due to the finite extension
in the fifth dimension. 

Since the global residual mass (\ref{eq:def_Mres_O}) depends on the observable $ {\cal O} $, 
it is necessary to determine the residual mass of the quark for any observable $ {\cal O} $.
So far, the residual mass has been only studied for the   
the pseudoscalar $ {\cal O}(y)= \bar q(y) \lambda^b \gamma_5 q(y) $, the pion interpolator.
It is interesting to see how the residual mass of the quark 
changes with respect to the physical observable.  
We have derived the generating functional for the $n$-point Green's function of fermion fields 
(\ref{eq:final_result_generating_functional}), 
which is essential for expressing the residual mass in terms of the quark propagator. 

For the observable $ {\cal O}(y)= \bar q(y) \lambda^b \gamma_5 q(y) $, 
we have obtained a new formula (\ref{eq:mres_Dc}) for the residual mass, 
which is useful in practice since it immediately gives the local residual mass 
once the 12 columns of the valence quark propagator $ (D_c + m_q)^{-1}(x,y) $ 
have been computed. For the global residual mass (\ref{eq:Mres_Dc}), 
it requires all-to-all quark propagators, or the low-lying eigenmodes of 
$ D = D_c ( 1 + r D_c)^{-1} $ for an estimation.

Moreover, we obtain the upper-bounds (\ref{eq:Mres_upper_bound_Q}) and (\ref{eq:Mres_upper_bound_G}) 
of the global residual mass, for one configuration as well as an ensemble of gauge configurations 
in lattice QCD with ODWF. 
They provide a guideline for designing lattice QCD simulation with ODWF. 
That is, with the input values of $ r=[2m_0(1-d m_0)]^{-1} $ and $ m_q $,    
how to choose the values of $ N_s $, $ \lambda_{min} $ and $ \lambda_{max} $ 
such that the residual mass meets the desired tolerance, versus the cost of the simulation. 

For the case when all eigenvalues of $ |H| $ fall inside the interval 
$ [\lambda_{min}, \lambda_{max}] $, only the second term in (\ref{eq:Mres_upper_bound_G}) 
contributes, then $ M_{res} \le d_Z/(2r) $, an exponentially decreasing function of $ N_s $.    
However, if there are some eigenvalues of $ |H| $ smaller than $ \lambda_{min} $, 
then the first term of (\ref{eq:Mres_upper_bound_G}) also contributes, which makes 
the exponential bound only hold for $ N_s < (N_s)^{thres} $, 
where $ (N_s)^{thres} $ depends on $ \lambda_{min} $ and the low-lying spectrum of $ |H| $.
Moreover, the first term of (\ref{eq:Mres_upper_bound_G})   
also shifts the exponential bound $ d_Z/(2r) $ to a larger value $ d_Z'/(2r) $,  
where the ratio $ d_Z'/d_Z \simeq 1-4 $, depending on $ m $, 
$\lambda_{min} $ and the low-lying spectrum of $ |H| $.
For $ N_s > (N_s)^{thres} $, the upper-bound would behave like $ 1/N_s^3 $, 
until it behaves like ``saturated" (decreasing slowly with respect to $ N_s $) 
at some larger $ N_s $.  

For ODWF with kernel $ H = H_w $, and plaquette gauge action with $ \beta = 5.95-6.10 $, 
then $ (N_s)^{thres} \simeq 16-20 $ for $ \lambda_{min} = 0.01-0.05 $ and $ \lambda_{max} = 6.20 $.
As demonstrated in Ref. \cite{Chiu:2011bm}, without fine tuning of $ \lambda_{min} $, 
the upper bound (\ref{eq:Mres_upper_bound}) gives a reliable estimate of 
$\langle m_{res} \rangle $ for $ N_s = 16 $, with the same order of magnitude. 

The existence of a range of $ N_s < (N_s)^{thres} $ for which the exponential bound $ d_Z'/(2r) $ 
holds is the salient feature of ODWF, which provides a viable way to 
preserve the chiral symmetry to a good precision (say, $ m_{res} a < 10^{-5} $) 
with a modest $ N_s $ (say, $ N_s \simeq 16 $).  

Finally, we have a few words about the efficiency of ODWF, in comparison with other variants of DWF.  
So far, the tests in Refs. \cite{Edwards:2005an,Brower:2012vk}
have been performed with quenched gauge configurations, by measuring 
$ \langle m_{res} \rangle $ versus the cost of computing the valence quark propagators.  
However, in full QCD, the cost/efficiency of HMC simulation also depends on the acceptance rate 
and the rate of topological tunnelling, which have not been addressed in Refs. \cite{Edwards:2005an,Brower:2012vk}.
We think it is premature to claim which lattice DWF is more efficient, only based on  
the cost of computing valence quark propagators versus the residual mass, without taking into account of   
the subtle issues (e.g., instability, topology freezing, etc.).

\begin{acknowledgments}

  This work is supported in part by the National Science Council
  (No. NSC99-2112-M-002-012-MY3) and NTU-CQSE (No. 10R80914-4).
  This paper was completed while TWC was 
  participating the workshop "New Frontiers in Lattice Gauge Theory" at  
  The Galileo Galilei Institute for Theoretical Physics (GGI) in Florence, Italy.
  TWC would like to thank GGI for the hospitality and the INFN for partial support 
  during the completion of this work.

\end{acknowledgments}

\end{document}